\documentclass[journal=jpccck]{achemso}

\usepackage[version=3]{mhchem} 
\usepackage{pdfpages}
\usepackage{textcomp}

\usepackage[colorlinks=true,citecolor=blue,urlcolor=blue]  {hyperref}

\def\textsubscript#1%
{$_{\text{#1}}$}

\usepackage{multirow}
\usepackage{array} 
\usepackage{makecell}

\usepackage{soul}

\newcommand{\tabincell}[2]{\begin{tabular}{@{}#1@{}}#2\end{tabular}}
\author{Yuefeng Yin}
\affiliation{Department of Materials Science and Engineering, Monash University, Clayton, Australia}
\author{Jiri Cervenka}
\affiliation{Department of Thin Films and Nanostructures, Institute of Physics ASCR, v. v. i., Prague, Czech Republic}
\author{Nikhil V. Medhekar}
\affiliation{Department of Materials Science and Engineering, Monash University, Clayton, Australia}
\email{nikhil.medhekar@monash.edu}

\title[title]
  {
Tunable Hybridization Between Electronic States of Graphene and Physisorbed Hexacene 
}

\abbreviations{IR,NMR,UV}
\keywords{graphene, electronic structure, acenes, external electric field, density functional theory}

\begin{document}
\begin{abstract}
\noindent Non-covalent functionalization via physisorption of organic molecules provides a scalable approach for modifying the electronic structure of graphene while preserving its excellent carrier mobilities. 
Here we investigated the physisorption of long-chain acenes, namely, hexacene and its fluorinated derivative perfluorohexacene, on bilayer graphene for tunable graphene devices using first principles methods. 
We find that the adsorption of these molecules leads to the formation of localized states in the electronic structure of graphene close to its Fermi level, which 
could be readily tuned by 
an external electric field in the range of $\pm$ 3 eV/nm. 
The electric field not only creates a variable band gap as large as 250 meV in bilayer graphene, but also strongly influences the charge redistribution within the molecule-graphene system. 
This charge redistribution 
is found to be weak enough not to induce strong surface doping, but strong enough to help preserve the electronic states near the Dirac point of graphene.
Our results further highlight graphene's potential for selective chemical sensing of physisorbed molecules under the external electric fields. 

\end{abstract}

\newpage

\subsection{Introduction}

Graphene---a planar layer of carbon atoms arranged in a hexagonal lattice---exhibits a linear electronic dispersion 
with the valence and conduction bands touching at the Dirac point\cite{Novoselov22102004}. 
As a result of this unique electronic structure, pristine graphene demonstrates an ultrahigh charge carrier mobility in excess of 200,000 cm\textsuperscript{2}V\textsuperscript{-1}s\textsuperscript{-1}, 
which can be 
exploited for novel, 
highly energy efficient electronic devices \cite{ADMA:ADMA201201587,ADMA:ADMA201290269,PhysRevLett.105.266601}. 
However, the development of graphene-based electronic devices is primarily 
hindered by the absence of an intrinsic band gap in its electronic structure\cite{novoselov2012roadmap}. 
Although various approaches for tailoring the electronic structure of graphene have been pursued in recent years\cite{C1CS15193B,doi:10.1021/ar3001487,C0JM02922J,zhang2011tailoring}, creating a significant band gap while maintaining large charge carrier mobilities in graphene 
remains a formidable challenge.

One strategy to modify the electronic structure of graphene is to utlilize  quantum confinement effects inherent in low dimensional structures such as  quasi one-dimensional graphene nanoribbons\cite{PhysRevLett.98.206805,wang2008room}. 
While this strategy can effectively induce 
the band gap, it also suffers from 
carrier scattering due to edge imperfections\cite{doi:10.1021/nl062132h}.
Another route is the chemical functionalization of graphene where the addition of covalent bonds to graphene (for example, via hydrogenation and fluorination) changes the hybridization of carbon atoms from $sp^2$ to $sp^3$\cite{doi:10.1080/00018732.2010.487978}. 
While such covalent functionalization successfully alters the electronic properties of graphene, it also leads to a severe degradation of its transport 
properties \cite{SMLL:SMLL201202196}. 
Recent demonstrations of heterostructures of graphene with other 2D materials (for example, boron nitride and transition metal dichalcogenides) also provide a possible option, but a consistent production of graphene heterostructure devices is difficult to control on large scale\cite{doi:10.1021/nn400280c,doi:10.1021/nl402062j}. 

Among the various approaches being pursued to modify the electronic structure of graphene, non-covalent functionalization via physisorption of organic molecules offers an interesting pathway\cite{zhang2011tailoring,C4NR06470D}. 
This approach relies on conserving the integrity of the $sp^2$-bonded carbon lattice and thus preserves the linear dispersion of electrons near the Dirac point\cite{C1CS15193B,C0JM02922J}. 
Moreover, the production of devices made of graphene with physisorbed molecules  can be readily assisted by molecular self-assembly and can  therefore be expected to be scalable.\cite{Mao2013132,Colson08042011}
Recent studies have suggested that in graphene physisorbed with small molecules such as NO$_\text{2}$ and NH$_\text{3}$, the application of a transverse external electric field can further enhance the tunability of the electronic structure of graphene by affecting the charge redistribution.\cite{zhang2009direct,tian2010band,zhang2011opening,doi:10.1021/jp212218w,:/content/aip/journal/jcp/134/4/10.1063/1.3541249,doi:10.1021/nl2025739}
Among the organic compounds that are amenable to physisorption on graphene, aromatic molecules are of particular interest \cite{PhysRevLett.102.135501,zhang2011tailoring}. 
The face-centered parallel stacking of aromatic molecules on graphene surface can lead to a stable hybrid system via van der Waals (vdW) interactions\cite{Colson08042011}, while the enhanced 
$\pi$-$\pi$ electron interaction is expected to influence the electronic structure of graphene\cite{zhang2011tailoring}. 
Moreover, addition of functional groups with high electron or hole affinity to the aromatic molecules has been suggested as an effective approach to induce strong charge doping in  graphene.\cite{chen2007surface,zhang2011opening,medina2011tuning,doi:10.1021/jp1107262,kozlov2011bandgap}. 
This can allow a vertical integration of graphene with physisorbed organic molecules with tunable transport characteristics such as charge injection barriers \cite{wehling2008molecular}.
However, recent reports indicate that a strong surface charge doping of graphene by molecules or electric field can cause a significant shift of the Fermi level into the valence or conduction band, and often lead to a severe deformation of the $\pi$ bands of graphene\cite{tian2010band,duong2012band}. 
As a result, the charge carrier mobility of graphene degrades, thus limiting the switching capability of graphene-based semiconducting devices\cite{duong2012band,doi:10.1021/jz4010174}. 
Therefore, identification of suitable organic molecules for non-covalent functionalization of graphene still remains 
an open challenge for controllable modification of its electronic properties.

Hexacene belongs to the group of acenes, the aromatic compounds formed by  linear fusion of benzene rings (C\textsubscript{4n+2}H\textsubscript{2n+4}). 
Long-chain acenes possess low-lying molecular orbitals that are expected to hybridize with $\pi$ electrons of graphene and thus influence its electronic structure.\cite{kadantsev2006electronic}. 
Furthermore, it has been shown that the edges of hexacene can be readily  functionalized with chemical groups with widely varying electron or hole affinity.\cite{doi:10.1021/ja051798v}. Recently, Watanabe {\em et al.} reported a successful way to synthesize hexacene that can remain stable up to 300~\textcelsius~in dark conditions\cite{watanabe2012synthesis}. 
Moreover, organic field effect transistors (OFET) devices made of hexacene have demonstrated a highest charge carrier mobility ever reported for organic semiconductors\cite{watanabe2012synthesis}. 
These observations suggest that hexacene can be an attractive candidate for  a stable physisorption on graphene. A good fundmental understanding of the 
electronic interactions between hexacene and graphene is therefore essential.

Here we systematically investigate the effect of physisorbed hexacene and perfluorohexacene (fluorinated hexacene) on the electronic properties of bilayer graphene using first principles density functional theory simulations. We use perfluorohexacene as an effective tool to down-shift the molecular energy levels relative to hexacene, and induce  significant $\pi$-$\pi^{\ast}$ interactions and symmetry breaking in bilayer graphene. 
We examine how the functional groups and adsorption geometry of molecules influence the stability and the electronic structure of the bilayer graphene-molecule system. 
We show that the adsorption of hexacene and perfluorohexacene on bilayer graphene leads to a significant charge redistribution and  the formation of localized states in graphene.
By applying an external electric field to bilayer graphene adsorbed with hexacene and perfluorohexacene, we demonstrate that the induced localized states in graphene can be effectively controlled, potentially providing a new strategy for graphene-based sensors for a selective sensing of weakly adsorbed molecules.


\subsection{Methods}
\label{simset}

\begin{figure}[btp]
\begin{center}
\includegraphics[scale=0.8]{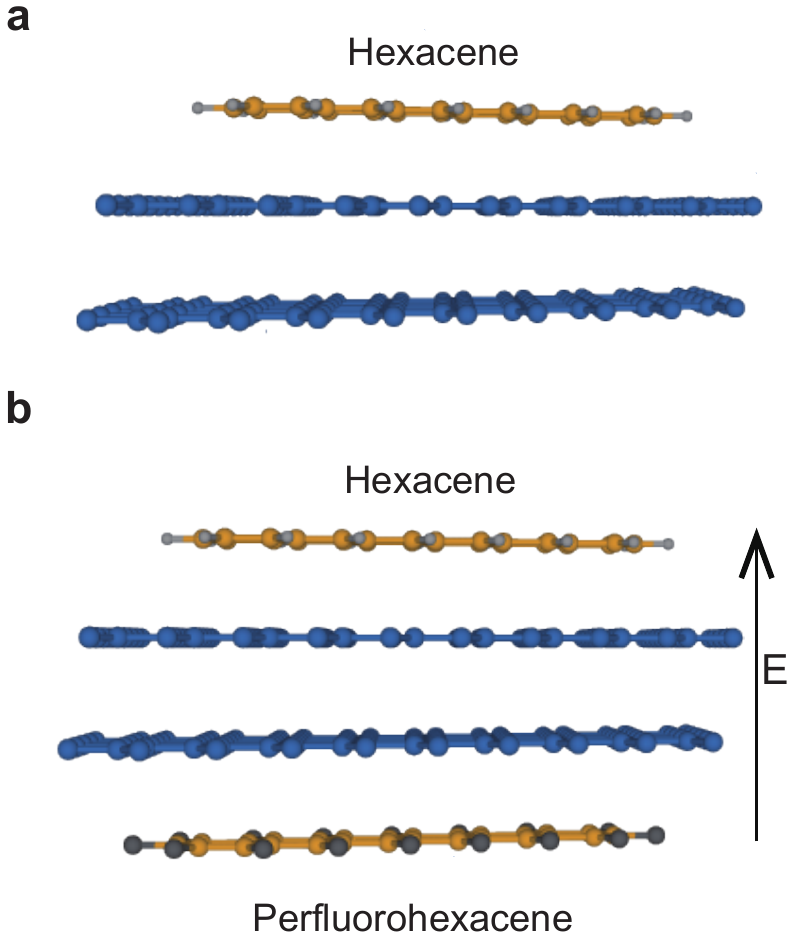}
\end{center}
\caption{\small A schematic illustrating (a) single- and  (b) dual-molecular adsorption of hexacene and perfluorohexacene on bilayer graphene. The arrow indicates the direction of an applied external electric field. 
}\label{fig1}
\end{figure}

To obtain optimized geometries and the electronic structures of all graphene-molecule systems considered in our study, we employed first principles  density functional theory as implemented in Vienna \emph{Ab Initio} Simulation Package \cite{kresse1996efficiency}.
We used the generalized gradient approximation of the Perdew-Burke-Ernzerhof form for the electron exchange-correlation functional\cite{PhysRevLett.77.3865}. 
The core and valence electrons were treated using projector augmented wave (PAW) scheme\cite{PhysRevB.59.1758} with a kinetic energy cut-off of 600 eV for the plane-wave basis set.
Since the generalized gradient approximation does not fully account for long-range dispersion interactions\cite{PhysRevB.81.081408}, we used a Grimme's semi-empirical functional\cite{JCC:JCC20495} to account for these 
interactions in the weakly bound graphene-molecule system. To benchmark the accuracy of this functional, we obtained the the equilibrium interlayer distance for pristine bilayer graphene to be 3.23 \AA, which is in good  agreement with the experimental values\cite{PhysRev.100.544}.
We used a periodic $8\times4$ graphene supercell for investigating the adsorption of hexacene and perfluorohexacene on Bernal-stacked bilayer graphene. 
A single molecule in this supercell represents a nearly monolayer coverage for  hexacene and perfluorohexacene on graphene with a molecular density of 9.846$\times$10\textsuperscript{-11} mol/cm\textsuperscript{2}.  
This magnitude of molecular density is representative of the reported coverage of the  aromatic molecules deposited on graphitic or graphene surfaces in various experimental studies\cite{PhysRevLett.52.2269,gotzen2010growth,lee2011surface}. 
In each case, the periodic images were separated by a 30 \AA \ vacuum, which was found to be large enough to avoid any spurious interactions between the  periodic images. 
 All structures were fully relaxed until the ionic forces  were smaller than 0.01 eV/\AA. Gaussian smearing was used for geometry relaxations, while  Bl\"{o}ch tetrahedral smearing was employed for subsequent calculations of electronic structures\cite{PhysRevB.49.16223}. 
Finally, for accurate calculations of the electronic structures, we used a fine  $6\times12\times1$ $\Gamma$-centred grid for sampling the Brillouin zone. 
 
First, we individually examined the adsorption of hexacene and perfluorohexacene on bilayer graphene using a single-molecular adsorption configuration as shown in Fig. \ref{fig1}(a).  
We considered two stacking sequences for this configuration, namely, $ABA$ and $ABC$, using the same notation as in the case of few-layer graphene\cite{lui2011observation}. 
We also considered the simultaneous adsorption of hexacene and perfluorohexacene in a dual-molecular  configuration shown in Fig. \ref{fig1}(b). 
Finally, the influence of external electric fields normal to the plane of graphene was investigated by introducing dipolar sheets at the center of supercells\cite{neugebauer1992adsorbate}.

\subsection{Results and Discussion}
\label{rd}

\noindent\textbf{Electronic Structure of Hexacene and Perfluorohexacene}

\begin{figure}[bt]
\begin{center}
\includegraphics[scale=0.8]{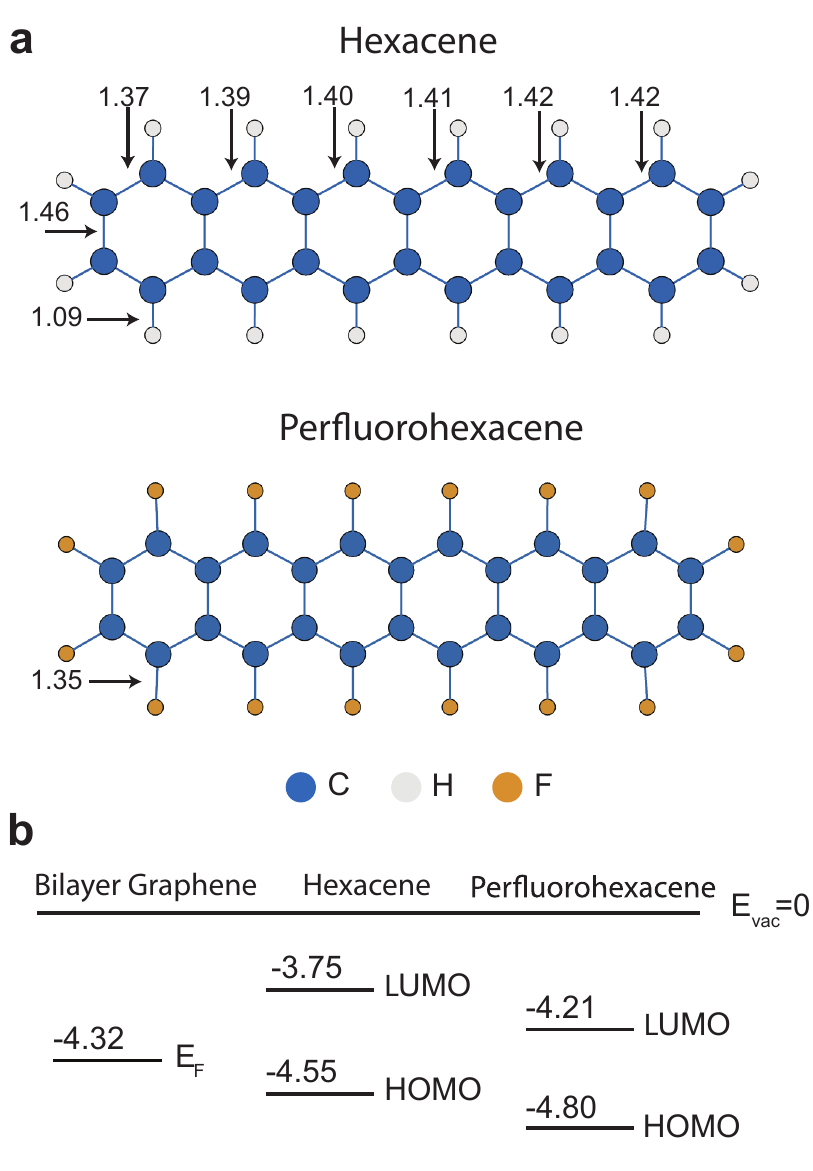}
\end{center}
\caption{\small (a) Relaxed geometries of hexacene and perfluorohexacene. The C-C, C-H and C-F bond lengths (in \AA) are indicated. (b) The energy level diagram of the band alignment between the Fermi level of bilayer graphene, and the HOMO and LUMO of hexacene and perfluorohexacene (in eV). The energy of vacuum is regarded as zero.}\label{fig2}
\end{figure}

\noindent We first obtained the optimized geometries of hexacene and perfluorohexacene molecules and calculated their electronic structure. The relaxed geometries of hexacene and perfluorohexacene are shown in Fig. \ref{fig2}(a). The average C-H bond length for hexacene and the average C-F bond length for perfluorohexacene were calculated to be 1.09 \AA \ and 1.35 \AA \ respectively. These bond lengths are in close agreement with previous structural calculations reported by Kadantsev et al.\cite{kadantsev2006electronic}. 
Figure \ref{fig2}(b) depicts an energy level diagram showing the highest occupied molecular orbital (HOMO) and the lowest unoccupied molecular orbital (LUMO) of hexacene and perfluorohexacene with respect to the position of the Fermi level of graphene. The HOMO-LUMO band gaps of hexacene and perfluorohexacene are narrow, namely, 0.80 eV and 0.59 eV, respectively. The calculated electronic band gap of these molecules is smaller than the experimentally determined band gaps by 0.5 eV  -- 0.6 eV due to a systematic underestimation of band gap values of semiconducting materials obtained by DFT calculations employing the generalized gradient approximation  \cite{watanabe2012synthesis,PhysRevLett.51.1884,PhysRevLett.51.1888}.

\begin{table*}[t]
\setlength{\abovedisplayskip}{0cm}
\caption{\small Adsorption energy and the average adsorption distance for monolayer coverage of hexacene and perfluorohexacene on bilayer graphene in single- and dual-molecular adsorption configurations. 
}
\begin{center}
\begin{tabular}{p{4.2cm}<{\centering}p{2.1cm}<{\centering}p{2.7cm}<{\centering}p{5.8cm}<{\centering}}
\hline
 & Stacking sequence & Adsorption energy (eV) & Average adsorption distance (\AA) \\
\hline
\multirow{2}{*}{Hexacene} & ABA &-1.756 & 3.19\\
&ABC& -1.751 & 3.19\\
\hline
\multirow{2}{*}{Perfluorohexacene} & ABA &-2.246 & 3.16\\
&ABC&-2.239&3.16\\
\hline
\multirow{4}{*}[1ex]{\tabincell{c}{Dual-molecular adosprion \\ (hexacene+perfluoroxacene)}}  &\multirow{2}*[1ex]{ABAB} &\multirow{2}*[1ex]{-4.015} & \tabincell{c}{3.20 (hexacene)     \\    3.17 (perfluorohexacene)} \\
&\multirow{2}{*}[1ex]{ABCA} &\multirow{2}{*}[1ex]{-4.005}&\tabincell{c}{3.19 (hexacene)\\3.16 (perfluorohexacene)}\\
\hline
\end{tabular}
\end{center}
\label{table4-1}
\end{table*}

\vspace{1cm}\noindent\textbf{Single- and Dual-Molecular Adsorption on Bilayer Graphene}

\noindent Next we obtained optimized ground state 
configurations of hexacene and pefluorohexacene adsorbed on bilayer graphene in single as well as dual-molecular configurations shown in Fig.\,\ref{fig1}. 
In each case, the adsorption energy is calculated as
$\Delta E=E_\text{{graphene/molecule}}-E_\text{{graphene}}-E_\text{{molecule}},$
where $E_\text{{graphene/molecule}}$ is the total energy of the fully-relaxed graphene-molecule supercell, while $E_\text{{graphene}}$ and $E_\text{{molecule}}$ is the energy of bilayer graphene and isolated molecules in the same supercell, respectively. 
A negative value of the adsorption energy indicates an exothermic, thermodynamically favorable adsorption. 
The values for adsorption energy and the average adsorption distance from the nearest graphene layer for each configuration are summarized in Table \,\ref{table4-1}. 
We find that the adsorption of both hexacene and pefluorohexacene on bilayer graphene is thermodynamically favorable in all  configurations considered. 
The resulting adsorption distance of the molecules from graphene is close to the interlayer distance between graphene layers ($\sim$3.23 \AA), indicating that graphene and the molecules are bound by weak van der Waals dispersion interactions. 
The adsorption distance and the adsorption energies are also nearly unaffected by the change of stacking sequence from ABC to ABA, also suggesting that the molecule-bilayer graphene interactions are largely confined between the molecule and the adjacent graphene layer.

Our results show that pefluorohexacene binds more strongly to bilayer graphene than hexacene (adsorption energy of -2.25 eV vs. -1.76 eV) in a single-molecular adsorption configuration. The stronger binding of pefluorohexacene correlates with a higher electron affinity and chemical reactivity of pefluorohexacene relative to hexacene. 
Finally, in the case of simultaneous, dual-molecular adsorption of hexacene and perfluorohexacene on bilayer graphene, the adsorption energy of the total hexacene and perfluorohexacene complex is nearly equal to the sum of the adsorption energies of individual molecules in single-molecular adsorption. This observation suggests that the adsorption system can be relatively easily adjusted from a single-molecular configuration to a dual-molecular configuration, or vice versa, at a very low energy cost. The adsorption distances of hexacene and perfluorohexacene in dual- molecular adsorption are found to be slightly larger than for the single-molecular adsorption cases.


\vspace{1cm}\noindent\textbf{Electronic Structure of Bilayer Graphene upon Adsorption}
\begin{figure}[t]
\begin{center}
\includegraphics[scale=0.9]{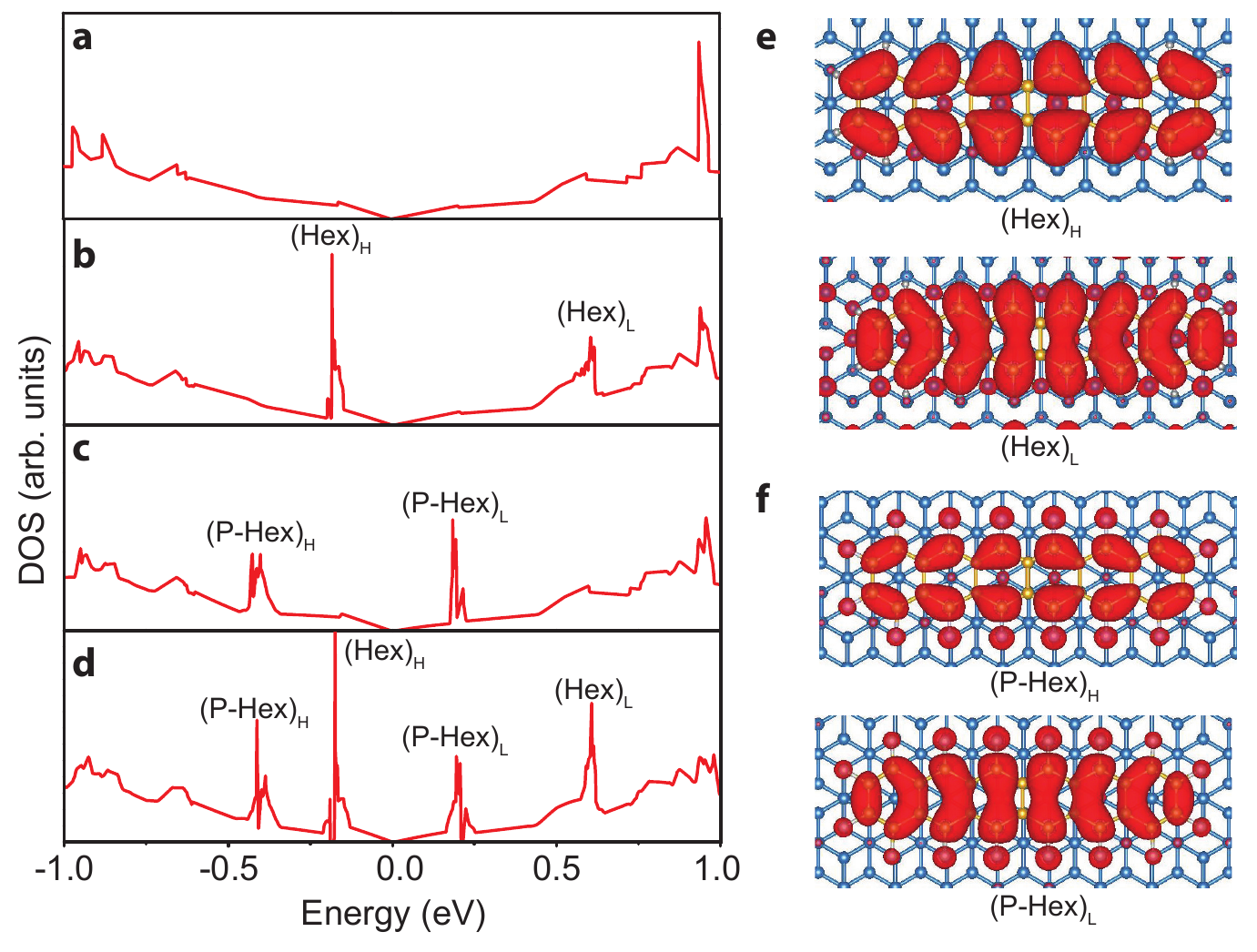}
\end{center}
\caption{\small (a) Total density of states of a pristine bilayer graphene. 
(b-d) Partial densities of states (PDOS) of bilayer graphene upon single-molecular adsorption of hexacene (b), single-molecular adsorption of perfluorohexacene (c), and   dual-molecular adsorption of hexacene and perfluorohexacene (d). 
In all cases, the molecular concentration is 9.85$\times$10\textsuperscript{-11} mol/cm\textsuperscript{2} and represents monolayer coverage. 
$(\text{Hex})_\text{H}$, $(\text{Hex})_\text{L}$, and $(\text{P-Hex})_\text{H}$, $(\text{P-Hex})_\text{L}$ denote the localized states induced by hybridization with HOMO and LUMO of hexacene, and HOMO and LUMO of perfluorohexacene, respectively. The Fermi level is set to zero in each case. 
(e-f) Partial charge density plots for the localized states in PDOS of bilayer graphene upon single-molecular adsorption of hexacene (e) and single-molecular adsorption of perfluorohexacene (f). The isosurface level is set to 0.0003 e/\AA\textsuperscript{3}.
}\label{fig4} 
\end{figure}

\noindent Following the structural optimisation, we next investigate the electronic properties of bilayer graphene upon molecular adsorption. 
Our results show that the molecular stacking sequence did not have any effect on the density of states  curves. Consequently, Fig.\,\ref{fig4} presents the partial density of states (PDOS) for bilayer graphene with single-molecular adsorption with ABA stacking and  for dual-molecular adsorption with ABCA stacking sequence.
We find that the physisorbed hexacene and perflurorohexacene has a negligible effect on the $\pi$ states of bilayer graphene in the vicinity of the Dirac point. No shift in the Fermi level of bilayer graphene was observed upon adsorption of hexacene or perfluorohexacene. 
The magnitude of the band gap induced in bilayer graphene upon adsorption of these molecules is also negligible (3 meV and 5 meV for hexacene and perflurorohexacene, respectively). These observations point to a weak interaction between graphene and the molecules. 
Nevertheless, the low lying highest occupied and lowest unoccupied molecular orbitals (HOMO and LUMO) of adsorbed molecules hybridize with $\pi/\pi^*$ states of graphene, giving rise to two localized states near the Fermi level of bilayer graphene (see Fig. \ref{fig4}(a-d)). These states were found in PDOS of bilayer graphene at -0.18 eV and 0.61 eV after the adsorption of hexacene and at -0.43 eV and 0.18 eV after the adsorption of perfluorohexacene. The position of HOMO and LUMO of hexacene and perfluorohexacene states is slightly altered in comparison to that of isolated molecules (refer Fig.\,\ref{fig2} (b)). 
To gain further insight into the localized states in graphene, we also plotted the corresponding partial charge densities by integrating the charge density in an energy range $\pm$0.02 eV around the localized peaks. The partial charge density plots (Fig. \ref{fig4}(e,f)) show the shape of HOMO/LUMO of hexacene/perfluohexacene, and present the signature of hybridization between graphene and the adsorbed molecules. It is evident that the induced localized states in graphene are located on the nearest carbon atoms of the top graphene layer.
The presence of these hybridized states near the Dirac point of graphene implies that both electrons and holes can be injected from graphene to molecules at a relative low energy cost. 
In dual-molecular adsorption configuration (Fig.\,\ref{fig4} (d)), the PDOS can be regarded as a superposition of the energy states from the single-molecular adsorption on graphene. Moreover, we observe that a 8 meV band gap is opened, equal to the sum of the band gap values of individual molecules. This indicates that the interaction between bilayer graphene and acene molecules in dual-molecular configuration is essentially governed by the interaction of a single graphene layer with the adjacent adsorbed molecule.

\begin{figure}[btp]
\begin{center}
\includegraphics[scale=1]{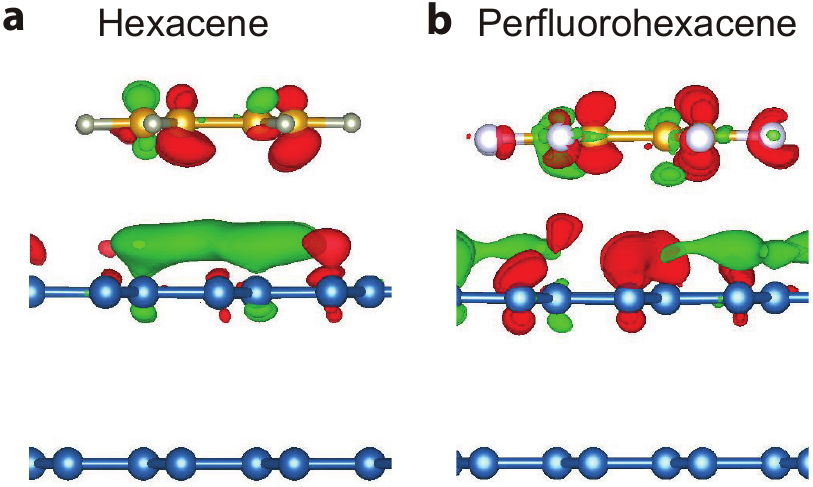}
\end{center}
\caption{\small The distribution of charge density difference for single-molecular adsorption of (a)  hexacene  and (b) perfluorohexacene  on bilayer graphene at monolayer coverage. Red and green isorufaces indicate the accumulation and depletion of electrons at a level of 0.0003 e/\AA\textsuperscript{3}, respectively. }\label{fig5}
\end{figure}

In order to further assess the influence of molecular adsorption on the electronic properties of graphene, we calculated the charge density difference as defined by
$\Delta\rho=\rho_\text{graphene/molecule}-\rho_\text{graphene}-\rho_\text{molecule},$
where $\rho_\text{graphene/molecule}$, $\rho_\text{graphene}$ and $\rho_\text{molecule}$ are the electronic charge densities of the adsorbed system, isolated graphene and the molecule,  respectively. With this definition, a positive value of $\Delta\rho$ indicates an accumulation of electronic charge and a negative value indicates a charge depletion. The distribution of the charge density difference for single-molecular adsorption of hexacene and perfluorohexacene is shown in Fig. \ref{fig5}.
In the case of adsorption of hexacene on bilayer graphene (Fig. \ref{fig5} (a)), the charges are depleted from the region 0.6 \AA --1.0 \AA~ above the top graphene layer and accumulated close to hexacene in the region 2.6 \AA -- 2.8 \AA \ above the graphene layer. 
This charge redistribution primarily arises from the electrostatic interaction between the aromatic rings of hexacene and graphene---the interaction  
between hydrogen atoms of hexacene and carbon atoms of graphene or hexacene is found to be negligible \cite{hsun2011electrostatic}. Overall, the interaction between hexacene and bilayer graphene is not strong enough to lead to a significant direct charge transfer between graphene and hexacene, but only to a charge redistribution on carbon atoms of the molecule and the nearest graphene layer.
In contrast to hexacene, the charge redistribution is significantly different for perflurorohexacene physisorbed on bilayer graphene as shown in Fig. \ref{fig5} (b). 
Since fluorine atoms are strong electron-attracting groups, the $\pi$ electrons are polarized away from the aromatic rings leading to  a relatively electron-deficient aromatic core of perfluorohexacene.
Similarly, a strong interaction between fluorine and graphene gives rise to a significant charge depletion from the carbon atoms in graphene close to fluorine atoms in perfluorohexacene, indicating a net charge transfer from graphene to perfluorohexacene. 
By comparing the charge distribution of hexacene and perfluorohexacene on bilayer graphene, it is evident that interaction between graphene and perfluorohexacene is largely controlled by the presence of fluorine functional groups.

\begin{figure}[t]
\begin{center}
\includegraphics[scale=0.9]{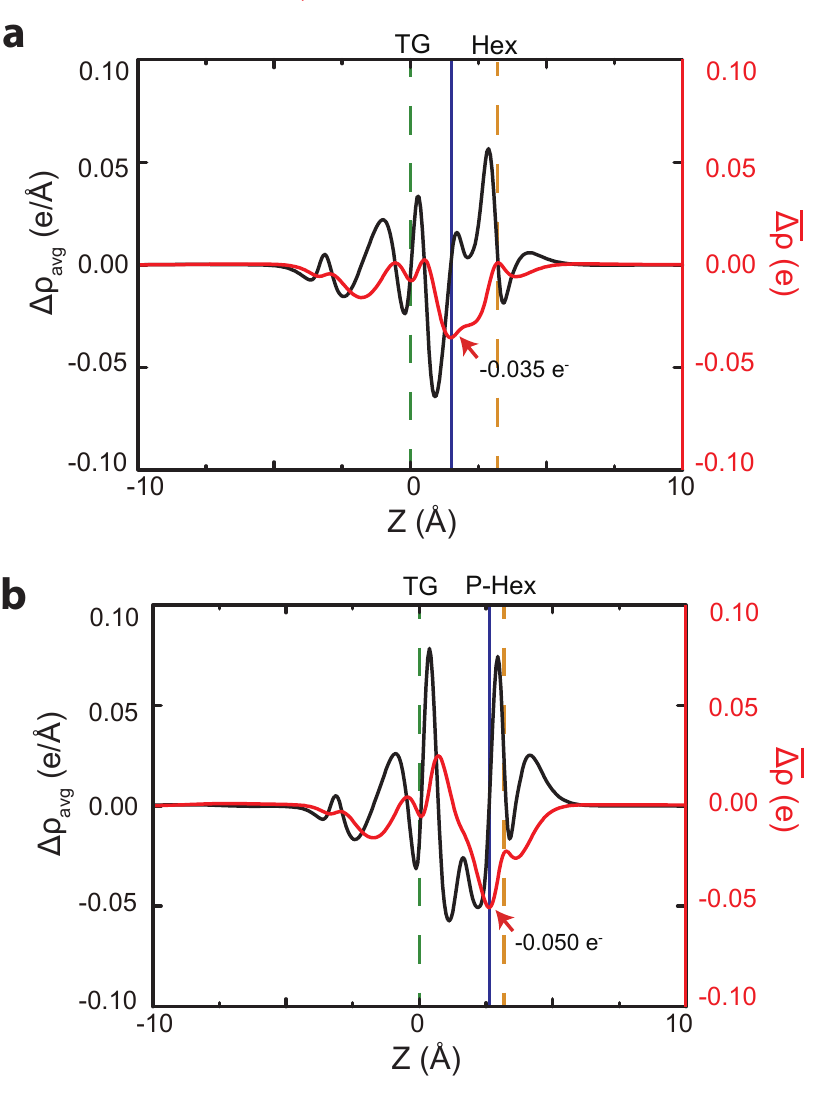}
\end{center}
\caption{\small The planar-averaged ($\Delta\rho_{\text{avg}}(z)$) and integrated planar-averaged ($\overline{\Delta\rho}(z)$) charge density difference curves for single-molecular adsorption of (a) hexacene and (b) perfluorohexacene on bilayer graphene at monolayer coverage. 
The positions of the top graphene layer (TG) and the molecule are marked by green and yellow dashed lines, respectively. Blue solid lines denote the neutral plane. 
}\label{fig6} 
\end{figure}

To quantify the charge transferred from graphene to adosprbed molecules, we plotted the variation of the planar-averaged and the integrated planar-averaged charge density difference as a function of the distance from the basal plane of graphene as shown in Fig. \,\ref{fig6}. 
The planar-averaged charge density difference $\Delta\rho_{\text{avg}}(z)$ along a plane parallel to the basal plane of graphene is obtained by integrating the charge density difference across the plane, whereas the integrated charge density difference $\overline{\Delta\rho}(z)$ is simply calculated by integrating $\Delta\rho_{\text{avg}}(z)$ curve from the boundary of the periodic box to the position of the plane. 
The extremum in the integrated charge density difference curve in the region between graphene and the adsorbed molecule (indicated by red lines in Fig. \,\ref{fig6}) denotes the neutral plane and the magnitude of the net charge transfer. 
According to this analysis,  graphene donates 0.035 $e^-$ per molecule to hexacene and a larger fraction 0.050 $e^-$ per molecule to perflurohexacene. 
The magnitude of net charge transfer obtained here is qualitatively consistent with Bader charge population analysis\cite{Henkelman2006354}, which yields a net charge transfer of 0.02  $e^-$ and 0.18  $e^-$ per molecule from graphene to hexacene and perfluorohexacene, respectively. 
The differences in the magnitudes of the net charge transfer obtained by  these two methods can be attributed to the different ways of calculating the net charges on each ion. 
Bader charge analysis takes into account both core charges and valence charges, while the charge density difference calculation is more suitable for interpreting the charge redistribution close to the Fermi level \cite{chan2008first}. Nevertheless, the small magnitude of charge transfer for both molecules is indicative of a weak interaction between the molecules and bilayer graphene.


Finally, few recent experimental studies have reported that as the density of the adsorbed acene molecules is  increased beyond the near-monolayer coverage considered here, the molecules tend to tilt rather than maintain a planar orientation on graphene \cite{lee2011surface,doi:10.1021/jp3103518}. 
In order to confirm this observation,
we have also studied the adorption of hexacene on bilayer graphene at a nominal molecular concentration of 1.31$\times$10\textsuperscript{-10} mol/cm\textsuperscript{2} using a $8\times3$ graphene supercell. 
Figure S1 in the Supporting Information shows the adsorption geometry as well as the electronic interaction between hexacene and bilayer graphene at this coverage. We find that hexacene shows a remarkable 11$^\circ$ tilt with respect to the basal plane of graphene, in qualitative agreement with the experimental observations \cite{lee2011surface,doi:10.1021/jp3103518}. 
This tilt arises due to a stronger repulsive intermolecular interaction between neighbouring hexacene molecules at high coverage. The cofacial $\pi$-$\pi$ interactions that contribute to the stability of hexacene adsorption at lower concentrations are disrupted, causing the adsorption at high coverage to be less energetically favorable (adsorption energy -1.534 eV per molecule). 
Moreover, the electronic interactions between graphene and hexacene also vary spatially, leading to an asymmetric charge redistribution pattern as shown in Fig. S1. 
Compared to the corresponding patterns at low coverages, a significantly larger charge rearrangement is observed in the region of the molecule where hexacene is closer to the graphene than in the region where hexacene is away from the graphene. 
This imbalance in the charge redistribution breaks the local symmetry of bilayer graphene, inducing a 54 meV band gap.

\vspace{1cm}\noindent{\textbf{Effect of the Applied External Electric Field}}

\begin{figure*}[t]
\begin{center}
\includegraphics[scale=0.9]{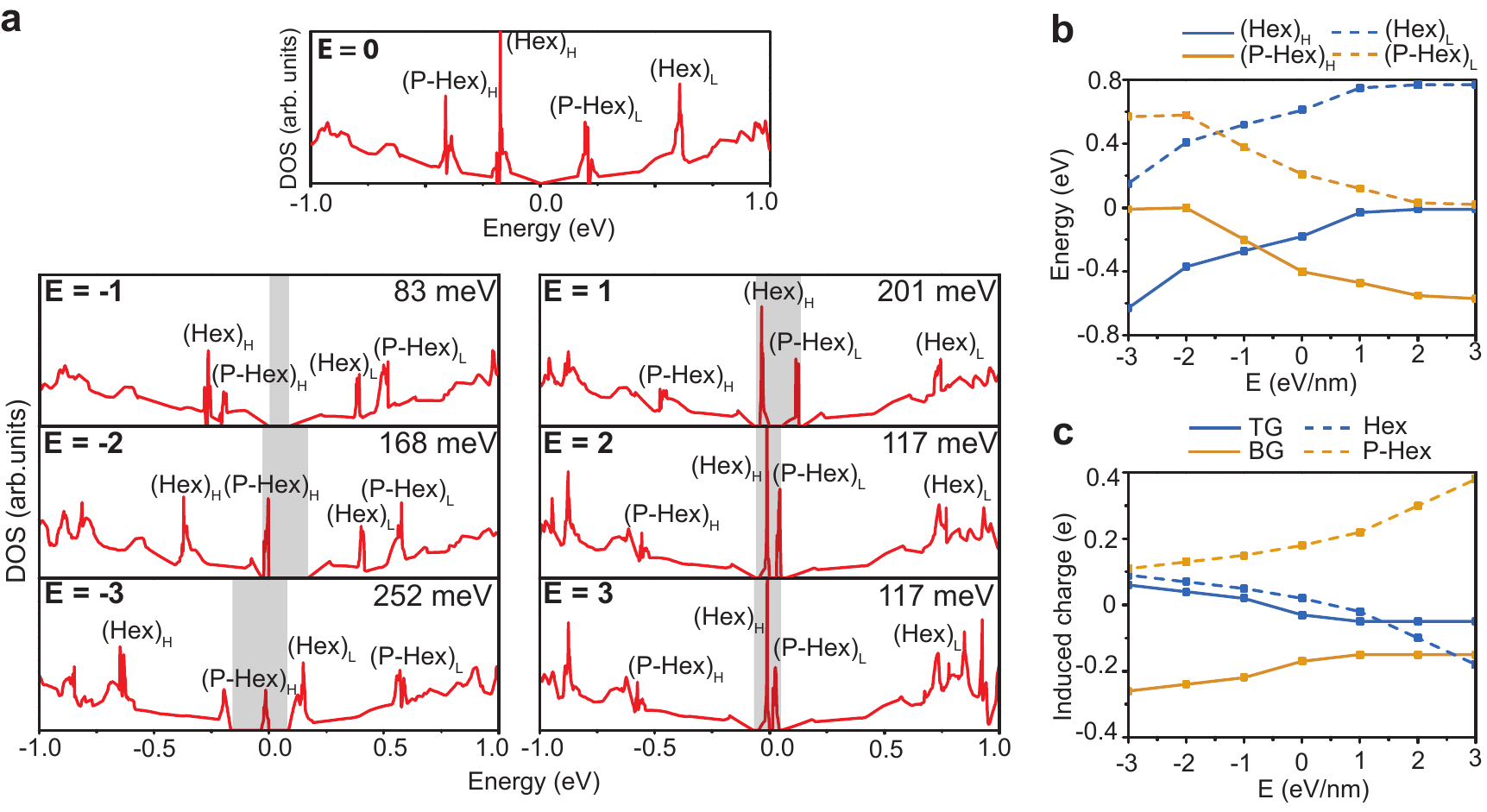}
\end{center}
\caption{\small Effect of external electric field on the electronic structure of bilayer graphene in dual-molecular adsorption of hexacene and perfluorohexacene at monolayer coverage. 
(a) PDOS of bilayer graphene as a function of external electrical field (in eV/nm). The peaks induced by hybridization with HOMO and LUMO states of hexacene and perfluorohexacene are denoted. The grey region indicates the band gap induced in bilayer graphene. The Fermi level is set to zero. 
(b) Positions of the localized states in bilayer graphene induced by the hybridization with HOMO and LUMO states of hexacene and perfluorohexacene. 
(c)  Induced charges in hexacene (Hex), perfluorohexacene (P-Hex), and top and bottom graphene layers (TG, BG), calculated by Bader charge population analysis.
}\label{fig7}
\end{figure*}

\begin{figure*}[tp]
\begin{center}
\includegraphics[scale=0.9]{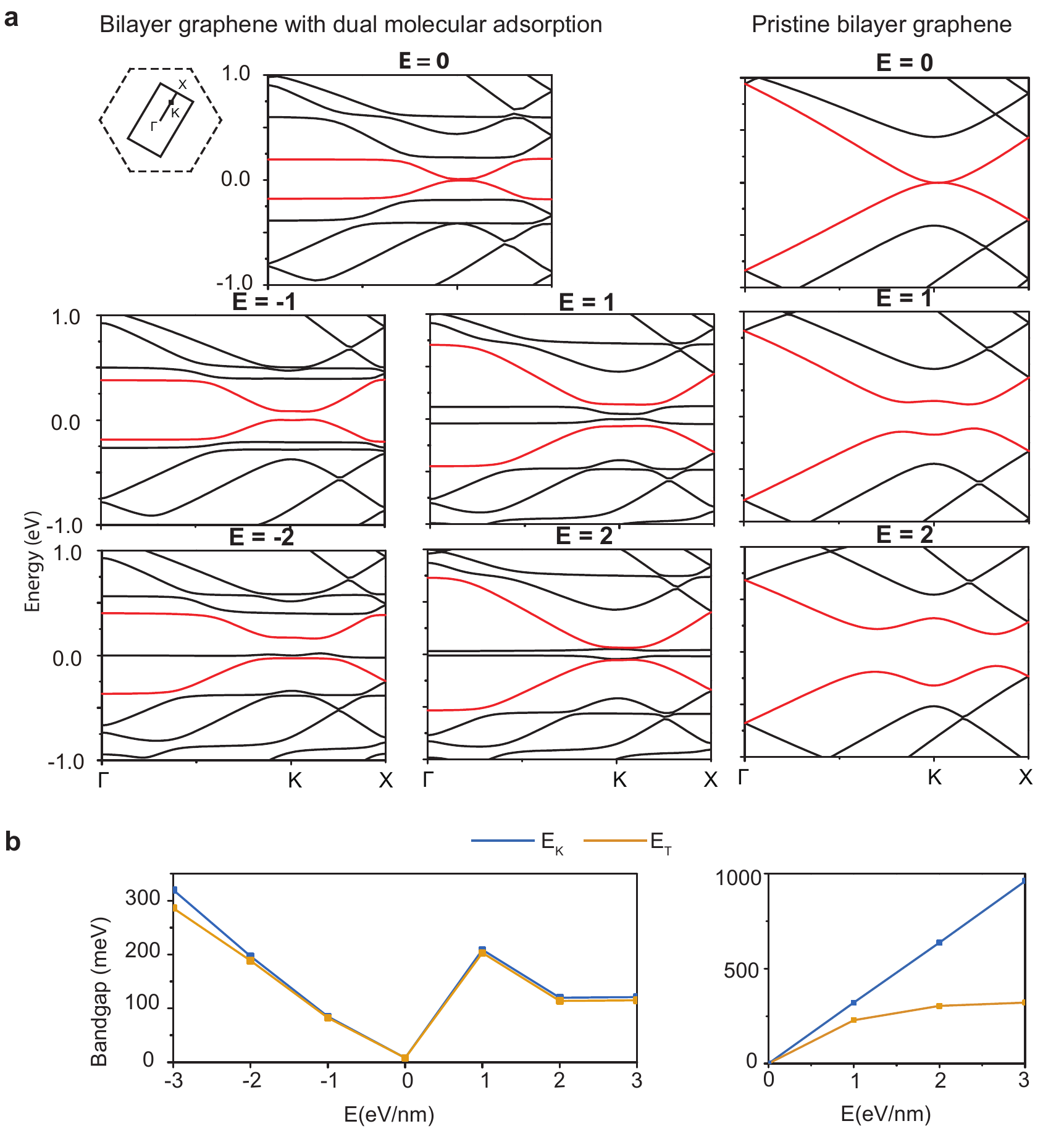}
\end{center}
\caption{\small (a) Electronic band structure diagrams for dual-molecular adsorption of hexacene and perfluorohexacene with monolayer coverage on bilayer graphene as a function of the strength of the electric field (in eV/nm). The corresponding diagrams for pristine bilayer graphene are also shown for comparison. The $\pi$ and $\pi^{\ast}$ bands of bilayer graphene are labeled in red color. The inset shows the first Brilloun zone of the supercell compared against that of the unit cell of graphene.
(b) The variation of the band gap at K (E\textsubscript{K}, blue) and the true band gap along $\Gamma$X line in the Brillouin zone (E\textsubscript{T}, yellow) with  the strength of the electric field.
}\label{fig9}
\end{figure*}

\noindent The results presented in earlier sections show that in general,  the adsorbed aromatic acene molecules interact weakly with bilayer graphene, leading to  the formation of localized states and a weak p-type doping of graphene. Since external electric fields can enhance the interactions between adsorbate and substrate\cite{duong2012band,doi:10.1021/jp212218w,:/content/aip/journal/jcp/134/4/10.1063/1.3541249}, next we investigate whether the external electric field could be effectively utilized to tune the electronic structure and molecule-specific localized states in bilayer graphene. 
Specifically, we studied the effect of the electric field in the range of -3 to 3 eV/nm, applied perpendicular to the basal plane of bilayer graphene in the dual-molecular  hexacene/bilayer graphene/perfluorohexacene adsorption configuration at a monolayer coverage as shown in Fig. \ref{fig1}(b). 
In our notation, a positive electric field is oriented towards hexacene from perfluorohexacene. 
Figure \ref{fig7}(a) presents PDOS of bilayer graphene as a function of the strength of the field.
The application of the electric field  leads to the opening of a considerable band gap  in bilayer graphene, as well as to the shift in the energy levels of the localized states arising  due to the hybridization with HOMO and LUMO states of the molecules. 
The external electric field causes an accumulation of electrons in one layer and a depletion of electrons in the other layer of bilayer graphene, thus breaking the interlayer symmetry \cite{zhang2009direct}.. 
The band gap generated in bilayer graphene 
is thus a result of the interplay between the field-induced interlayer symmetry breaking and the asymmetric charge transfer between graphene and the adsorbed hexacene and perfluorohexacene molecules. 
In several cases, the magnitude of band gap is difficult to determine from the PDOS of bilayer graphene alone as the gap region is occupied by the localized states. 
In order to correctly identify the band gaps, we have calculated the electronic band structure for each case as shown in  Fig. \ref{fig9}. 
The  localized states can then be readily identified from these band structure diagrams  as the flat bands between the $\pi$ and $\pi^{\ast}$ bands of bilayer graphene.
   
It is evident that both the shape of the PDOS as well as the magnitude of the band gap in bilayer PDOS shows a strong dependence on the strength and the direction of the external electric field. 
For example, the band gap increases linearly with the electric field till a maximum of $\sim$250 meV  when the field is oriented from hexacene towards perfluorohexacene (that is, a negative electric field). 
For a positive field, the band gap reaches $\sim$200 meV for the field magnitude of 1 eV/nm, then reduces to $\sim$115 meV for the field of 2 eV/nm or greater. 
This observed contrast in the trend between negative and positive fields can be attributed to the distinct charge transfer behavior of the adsorbed hexacene and perfluorohexacene molecules as shown in Fig. \ref{fig7} (b). 
For instance, for fields $E \geq 2$ eV/nm, the HOMO of perfluorohexacene and the LUMO of hexacene are pinned in the vicinity of the Fermi level. 
The pinning of these localized states near the Fermi level enhances the driving force for transferring charge from hexacene to perfluorohexacene, which is favored by the application of a positive electric field. 
Therefore the charge transfer between hexacene and perfluorohexacene is maximized as shown in Fig. \ref{fig7} (c). 
The charge inequivalence between two graphene layers is reduced for positive electric fields greater than 1 eV/nm, 
leading to a saturation of the band gap. 
For negative fields on the other hand, the charge transfer trend is reversed. 
The energy difference between the HOMO of perfluorohexacene and the LUMO of hexacene increases with the strength of the negative electric field. 
Therefore the charge transfer between the two graphene layers is less affected and the magnitude of band gap increases rapidly with the magnitude of the field.

Figure \ref{fig9}(a) also shows a comparison between the electronic band structure diagrams of bilayer graphene adsorbed with hexacene and perfluorohexacene to those with the pristine bilayer graphene.
Our obtained band structures for pristine graphene are in good agreement with previous works \cite{doi:10.1021/nl1039499}. 
It is evident that due to the deformation of $\pi$ and $\pi^{\ast}$ bands, the field-induced band gap in pristine bilayer graphene is no longer located at the K point, but instead  along $\Gamma$-X line of the Brillouin zone. 
However, the $\pi$ and $\pi^{\ast}$ bands of bilayer graphene with dual-molecular adsorption are less deformed close to the Fermi level due to the screening by molecules, indicating that the energy dispersion relationship in the vicinity of the K point is relatively well preserved. 
This can be clearly seen by comparing the magnitude of band gap at K (E\textsubscript{K}) and the true band gap along $\Gamma$-X line (E\textsubscript{T}) presented in Fig. \ref{fig9}(b). 
It can be seen that in general, the range of band gaps that can be induced in  bilayer graphene with molecular adsorption (100 meV -- 250 meV) is comparable to the  pristine bilayer graphene (200 meV -- 300 meV). 
Overall, the band structures follow a similar trend as the DOS plots of dual molecular adsorption on bilayer graphene shown  in Fig. \ref{fig7} (a), where graphene  $\pi$ and $\pi^{\ast}$ states hybridize with the localized HOMO and LUMO bands of the molecules, forming new localized bands with a flat dispersion. 
These results show that the energy of these localized states can be varied as a function of the strength and polarity of the applied electric field, thereby modifying the  electronic structure of bilayer graphene. 
Finally, for completeness, we briefly compare the  electronic structure of bilayer graphene with dual-molecular adsorption under the electric field to that of  monolayer graphene. Figure S2 in the Supporting Information shows band structure diagrams for dual-molecular adsorption of hexacene and perfluorohexacene on monolayer graphene as a function of the strength of the electric field. We find that the electronic structure of monolayer graphene with dual-molecular adsorption is significantly different from the bilayer graphene for negative electric fields. The  localized states induced in monolayer graphene due to hybidization with the HOMO/LUMO states of the adsorbed molecules are relatively further away from the Fermi level of graphene. Moreover, negative electric fields lead to much lower band gaps in monolayer graphene (less than 50 meV). For positive electric fields, both monolayer and bilayer graphene show a similar magnitude of the band gaps ($\sim$120 meV). 
When compared with Fig. \ref{fig9}, these observations highlight the interplay between the charge transfer  and the breaking of the symmetry between the top and bottom layer in bilayer graphene. In general, these results suggest that by the application of external electric field to bilayer graphene with dual molecular gating, the electronic structure can be more flexibly controlled, leading to opening of considerable band gaps.

\subsection{Conclusions}

Here we have used self-consistent density functional theory calculations to study the effect of physisorption of hexacene and its fluorine derivative, perfluorohexacene, on the electronic structure of bilayer graphene. 
We find that although the overall interaction between graphene and molecules is weak, the adsorption of these molecules results in a significant charge redistribution. 
This charge redistribution gives rise to the hybridization of HOMO/LUMO energy levels of the molecules with the $\pi$ electrons of graphene, leading to the formation of localized states in bilayer graphene. 
We have shown that the external electric fields can be used to tune the electronic properties of graphene-molecule system, effectively opening large band gaps of the order of 250 meV in bilayer graphene. 
Furthermore, external electric fields can also infleunce the energies of the localized states in graphene, an effect that can be utilized in organic field effect transistor (OFET) devices by aligning the electronic states of acene channels with that of graphene electrodes. This effect can also be  potentially useful in the sensing of different organic molecules on the surface of graphene transistors. Graphene transistors have proven to be extremely sensitive sensors\cite{schedin2007detection}, but their selectivity remains a major problem for their practical use. 
In summary, we have shown that hexacene, a stable and high mobility organic electronic material, and its derivatives, are promising candidates for surface electronic structure modification of graphene for potential applications in organic electronics and sensing.

\subsection{Acknowledgement}
Authors gratefully acknowledge computational support from Monash Sun Grid as well as iVEC and NCI national computing facilities. 

\subsection{Supporting Information}
Additional figure for the adsorption of hexacene on bilayer graphene in the 8$\times$3 graphene supercell and band structure diagrams for dual-molecular adsorption of hexacene and perfluorohexacene on monolayer graphene.


\bibliography{ref}


\includepdf[pages={1}]{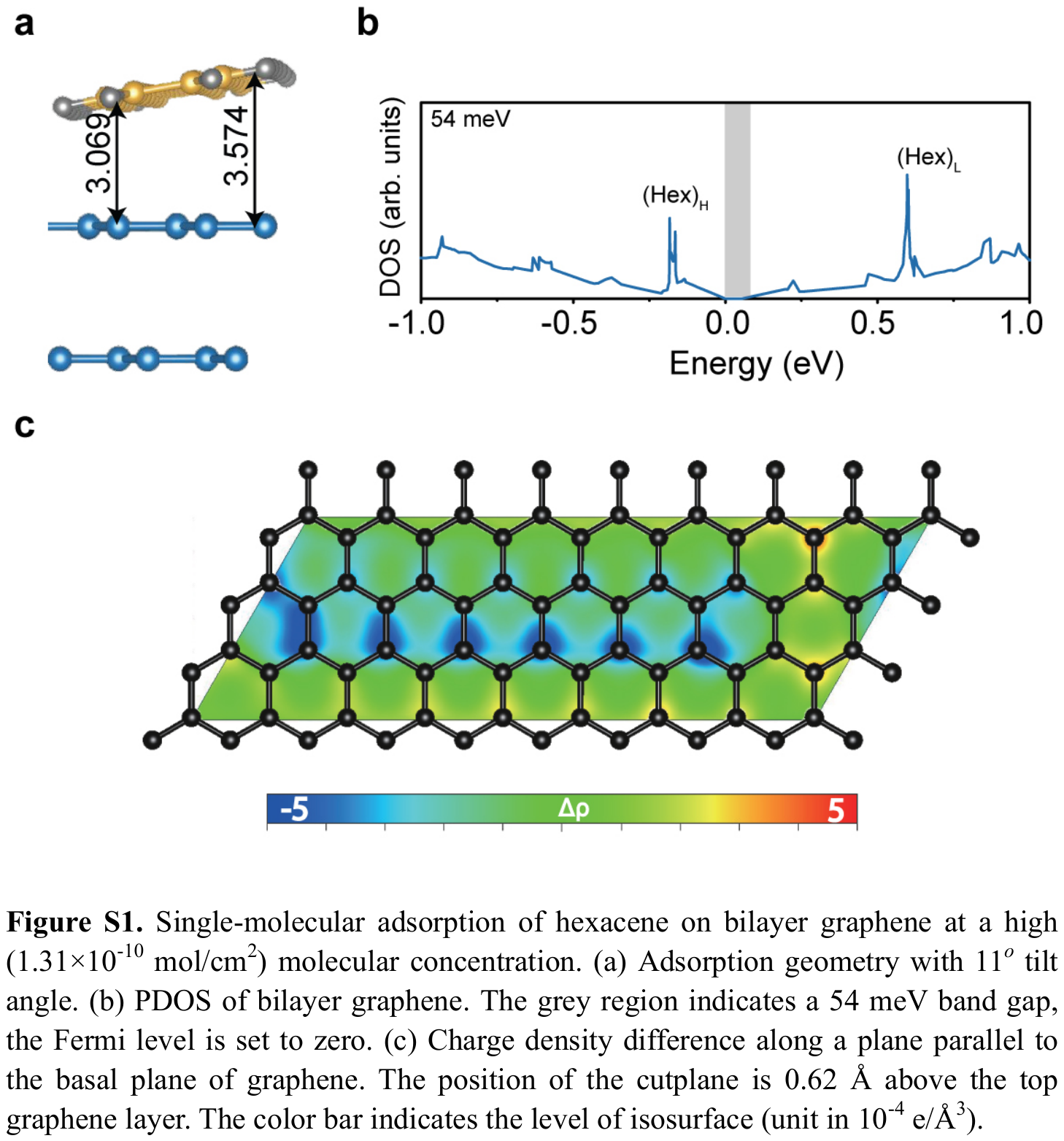}
\includepdf[pages={2}]{SI.pdf}

%
%
%
%
%





\end{document}